# Plasmonic Bragg Reflectors for Subwavelength Hole Arrays in a Metal Film

Pramodha Marthandam and Reuven Gordon, *Member, IEEE*

*Abstract*— We present plasmonic Bragg reflectors for increasing the extraordinary optical transmission through periodic arrays of subwavelength apertures in a metal film. The increase in transmission arises by preventing loss at the edges of the array. The plasmonic Bragg reflectors have periodicity of half the usual array's periodicity. Nano-hole arrays flanked thus by reflectors exhibited double the transmission as compared to the arrays without the reflectors. These structures may also be used to isolate devices in plasmonic integrated circuits.

## I. INTRODUCTION

SUBWAVELENGTH apertures in thin metal films show extraordinary optical transmission properties with resonance transmission wavelengths being dictated by the periodicity of the array [1]. Strong enhancements in the intensity of light transmitted through an isolated aperture, as well as improved directionality of the transmitted beam have been achieved by placing grooves around the aperture [2],[3]. Transmission enhancement by shallow grooves around a single subwavelength slit depends on groove periodicity [4]. The excited SP modes of nano-hole arrays have been found to scatter off the edges of the array and propagate coherently up to distances of several microns [5]. Since SP scattering off the edges of arrays constitute a loss mechanism, recapturing the scattered waves can be achieved by the use of layers of reflectors with a periodicity of $\lambda/2$, where $\lambda$ is the periodicity of the nano-hole array. These plasmonic Bragg reflectors (PBRs) also have the benefit of isolating one array from another, thereby preventing cross-talk between the arrays.

In this paper, we demonstrate experimentally the use of PBRs to enhanced the extraordinary optical transmission of subwavelength hole arrays. We show an increase of nearly double the extraordinary transmission for shallow groove PBRs. This may be used in the future to effectively isolate plasmonic devices in the context of plasmonic integrated circuits.

## II. SAMPLE FABRICATION AND TRANSMISSION MEASUREMENTS

### A. Fabrication of arrays

Figure 1 shows scanning electron micrographs of three arrays without PBRs, with dimpled PBRs and with groove PBRs, labeled A, B and C. The subwavelength hole arrays were fabricated on a 100nm thick gold film on a glass substrate by focused ion beam milling. The Gallium beam milling was done at 30 kV with spot size of 7 nm. Each array took approximately 2-3 minutes to mill. A 5 nm thick chromium film provides adhesion between the gold and the glass. Three arrays were fabricated, viz. array A: 15 μm × 30 μm arrays of holes of diameter 150 nm and periodicity 600nm with the long edge of the array oriented vertically, array B: same as array A with 7.5 μm × 30 μm arrays of circular dimples of diameter 150 nm and periodicity 300 nm flanking the right and left hand side of the holes, and array C: same as Array A with 7.5 μm×30 μm arrays of rectangular grooves of width 75 nm flanking the right and left hand side of the holes. The approximate milling depth of the grooves and the dimples was 35 nm.

### B. Transmission measurements

The sample was illuminated by a broadband halogen source focused down on the arrays by a 40× microscope objective. The spot size was approximately 35 μm. The incident light was polarized normal to the plane of the PBRs. The transmitted light was collected by a 200 μm core fiber. The incident and collected light were both normal to the surface of the sample. Transmitted spectra were measured by UV-Vis

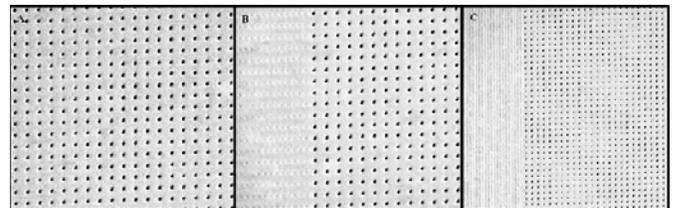

Fig. 1. Scanning electron micrographs of the fabricated arrays. Array A: 15 μm × 30 μm array of 150 nm holes, and periodicity 600 nm, Array B: same as A flanked with 7.5 μm × 30 μm array of 150 nm dimples with periodicity 300 nm, Array C: same as A flanked with 7.5 μm × 30 μm array of 75 nm grooves with periodicity 300 nm

Manuscript received June 27, 2007.

Pramodha Marthandam is with the Department of Electrical and Computing Engineering, University of Victoria, P.O. Box 3055 STN CSS, Victoria, BC, V8W 3P6

Reuven Gordon is with the Department of Electrical and Computing Engineering, University of Victoria, P.O. Box 3055 STN CSS, Victoria, BC, V8W 3P6 (phone: 250-472-5179; fax: 250-721-6052; e-mail: rgordon@uvic.ca).

The authors acknowledge support from CIPI, NSERC (Discovery Grant and Research Tools and Instruments) and CFI/BCKDF.

spectrometer. An integration time of 1 second was used, and 10 scan averages were collected to reduce noise.

## III. RESULTS AND DISCUSSION

Figure 2 shows the transmission spectra of the three arrays A, B and C. The arrays have a transmission resonance at a wavelength of 691 nm. The enhancement in the transmission in arrays B and C due to the PBRs is evident from the transmission spectra. At the resonant wavelength, the transmission is enhanced by 1.4 times for array B, the array flanked by circular dimples, and 1.9 times for array C, the array with shallow groove reflectors.

The grooves and dimples, by virtue of having a periodicity and being placed from the array boundary at a distance of $\lambda/2$, reflect the light back into the array. The reflected light then constructively interferes with the light actually transmitted by the apertures in the array, thus increasing the net transmission. It is clear that the enhanced transmission is not due to direct transmission through grooves or dimples in the PBRs, because the optical spectrum maintains the resonance features of enhanced optical transmission. Further experiments are underway to quantify the ability of the PBR to isolate one array from another.

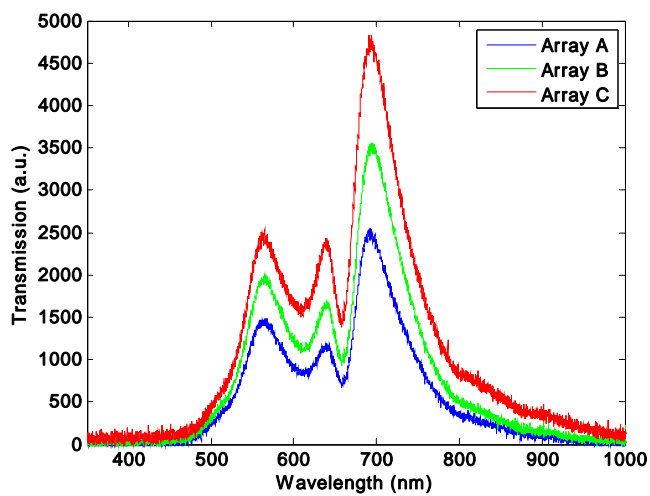

Fig. 2. Transmission spectra of arrays A, B and C. Transmission is enhanced by 1.9 times in Array C, and by 1.4 times in Array B.

## IV. CONCLUSIONS

PBRs were used to increase the extraordinary transmission of light through subwavelength hole arrays. The PBRs were made up of partially milled dimples and grooves at half period of the subwavelength hole array. The PBRs worked by recapturing light scattered out of the boundaries of the arrays and thereby reduced the scattering loss and increased the overall transmission. Variations in the number of reflector layers as well as positioning of the reflectors with respect to the array will be studied in the near future. The PBR may be used to isolate plasmonic devices in the context of plasmonic integrated circuits or optical filters. Further work is underway to investigate their ability to isolate separate arrays by preventing coupling between them from the propagation of scattered SPs.

ACKNOWLEDGMENT

The authors thank K.L. Kavanagh for useful discussions and the use of the NanoImaging Facility at SFU.